\documentclass[letterpaper,english,reprint, aps]{revtex4-1}
\pdfoutput=1
\usepackage[T1]{fontenc}
\usepackage[utf8]{inputenc}
\usepackage{babel}
\usepackage{verbatim}
\usepackage{float}
\usepackage{amsmath}
\usepackage{amssymb}
\usepackage{graphicx}
\usepackage{booktabs}
\usepackage{graphicx}
\usepackage{subscript}
\usepackage[unicode=true,pdfusetitle,
 bookmarks=true,bookmarksnumbered=false,bookmarksopen=false,
 breaklinks=false,pdfborder={0 0 1},backref=false,colorlinks=false]
 {hyperref}

\makeatletter

\newcommand{\lyxmathsym}[1]{\ifmmode\begingroup\def\b@ld{bold}
  \text{\ifx\math@version\b@ld\bfseries\fi#1}\endgroup\else#1\fi}

\usepackage{geometry} 
\usepackage{amsmath}
\usepackage{xparse}

\newcommand{\ket}[1]{\ensuremath{\left|#1\right\rangle}}

\newmuskip\pFqmuskip
\newcommand*\pFq[6][8]{%
  \begingroup 
  \pFqmuskip=#1mu\relax
  \mathchardef\normalcomma=\mathcode`,
  \mathcode`\,=\string"8000
  \begingroup\lccode`\~=`\,
  \lowercase{\endgroup\let~}\pFqcomma
  {}_{#2}F_{#3}{\left[\genfrac..{0pt}{}{#4}{#5};#6\right]}%
  \endgroup
}
\newcommand{\pFqcomma}{{\normalcomma}\mskip\pFqmuskip}

\ExplSyntaxOn
\NewDocumentCommand{\MeijerG}{smmmm}
 {
  \IfBooleanTF{#1}
   {
    \vic_meijerg:nnnnnn { #2 } { #3 } { #4 } { #5 } { small } { }
   }
   {
    \vic_meijerg:nnnnnn { #2 } { #3 } { #4 } { #5 } { } { \; }
   }
 }

\seq_new:N \l__vic_meijerg_args_in_seq
\seq_new:N \l__vic_meijerg_args_out_seq

\cs_new_protected:Nn \vic_meijerg:nnnnnn
 {
  \seq_set_split:Nnn \l__vic_meijerg_args_in_seq { | } { #3 }
  \seq_clear:N \l__vic_meijerg_args_out_seq  
  \seq_map_inline:Nn \l__vic_meijerg_args_in_seq
   {
    \seq_put_right:Nn \l__vic_meijerg_args_out_seq
     {
      \begin{#5matrix} ##1 \end{#5matrix}
     }
   }
  G\sp{#1}\sb{#2}
  \left(
  \seq_use:Nn \l__vic_meijerg_args_out_seq { #6\middle|#6 }
  #6\middle|#6
  #4
  \right)
 }
\ExplSyntaxOff

\makeatother

\begin{document}
\title{Tunable Excitons in Rhombohedral Trilayer Graphene}
\author{M. F. C. Martins Quintela$^{1,2}$, N. M. R. Peres$^{1,2}$}
\address{$^{1}$Department and Centre of Physics, University
	of Minho, Campus of Gualtar, 4710-057, Braga, Portugal}
\address{$^{2}$International Iberian Nanotechnology Laboratory (INL), Av. Mestre
	Jos{\'e} Veiga, 4715-330, Braga, Portugal}
\begin{abstract}
	Trilayer graphene is receiving an increasing level of attention due to its stacking--dependent magnetoelectric and optoelectric properties, and its more robust ferromagnetism relative to monolayer and bilayer variants.
	Additionally, rhombohedral stacked trilayer graphene presents the possibility of easily opening a gap via either an external electric field perpendicular to the layers, or via the application of external strain. 
	In this paper, we consider an external electric field to open a bandgap in rhombohedral trilayer graphene and study the excitonic optical response of the system. 
	This is done via the combination of a tight binding model with the Bethe--Salpeter equation, solved semi--analytically and requiring only a simple numerical quadrature. 
	We then discuss the valley--dependent optical selection rules, followed by the computation of the excitonic linear optical conductivity for the case of a rhombohedral graphene trilayer encapsulated in hexagonal boron nitride. 
	The tunability of the excitonic resonances via an external field is also discussed, together with the increasing localization of the excitonic states as the field increases. 
\end{abstract}
\maketitle

\section{Introduction}
Ever since the discovery and isolation of graphene\cite{doi:10.1126/science.1102896}, a plethora of different layered materials have been studied in detail. Of these layered materials, we specifically mention hexagonal boron nitride (hBN)\cite{Caldwell2019} and transition metal dichalcogenides (TMDs)\cite{RevModPhys.90.021001}. In hBN, the large bandgap and strong second--order nonlinearities make it well--suited for deep--UV optoelectronics\cite{Caldwell2019,Kubota2007}. Regarding TMDs, these  display strong spin--orbit coupling and breaking of inversion symmetry, leading to coupled spin and valley physics, and valley--selective optical excitations \cite{schneider_two-dimensional_2018,hsu_dielectric_2019,zhang_magnetic_2017,PhysRevLett.108.196802}.

The optical response of these materials is dominated by excitons\cite{nwu078}, which consist of bound electron--hole pairs. These are created by the excitation of an electron from the valence band to the conduction band, leaving behind a hole in the valence band. The electrostatic interaction\cite{rytova1967,keldysh1979coulomb} between the pair leads to the formation of a bound state inside the bandgap of the material, forming a Hydrogen--like system. The large binding energies of excitons, together with their efficient coupling with light, makes them a highly relevant and a rich field of research. Recently, various works have focused their attention on the optical response of excitons in TMDs, both in the linear regime \cite{Merkl2019,henriques2020optical} as well as in the non--linear regime \cite{henriques2021calculation,PhysRevB.104.205433}.

As graphene lacks the necessary bandgap for the formation of electron--hole bound states, excitonic phenomena are absent in pristine graphene monolayers. Graphene multilayers can, however, be engineered to present a bangap and, as such, host bound electron--hole pairs. A simple example of this is biased bilayer graphene, where a external perpendicular electric field is applied to a pair of stacked graphene monolayers, opening a tunable bandgap and allowing the formation of excitons. This system, encapsulated in hBN, was the subject of recent experimental \cite{doi:10.1126/science.aam9175} and theoretical \cite{doi:10.1021/nl902932k,PhysRevB.105.045411,sauer2021exciton} studies. 

A less studied system is that of biased trilayer graphene, where three graphene monolayers are stacked and an external perpendicular electric field is applied to the multilayer. 
Ferromagnetism has been shown to be more robust in trilayer graphene than in either monolayer and bilayer graphene, specifically when the layers are stacked in an ABC fashion (\emph{i.e.}, rhombohedral stacking) rather than in an ABA fashion (\emph{i.e.}, Bernal stacking)
\cite{PhysRevB.87.115414,PhysRevB.101.245411}. While the largest bandgaps obtained in bilayer graphene systems have been around hundreds of $\mathrm{meV}$, in trilayer graphene bandgaps of around $2\,\mathrm{eV}$ have been obtained by tuning the interlayer coupling via compression of a few $\mathrm{GPa}$\cite{Ke9186}. 

Recent experimental and theoretical works have also shown that several transport properties depend on the stacking order, including but not limited to thermoelectric\cite{PhysRevB.86.115414} and magnetoelectric\cite{PhysRevB.84.161408} transport. Additionally, it has been shown that a considerable gap can be opened in ABC--stacked trilayer graphene via an external electric field, while the same does not occur in ABA--stacked trilayer graphene under the same situations\cite{Lui2011,Rashidian_2014}. 
The possibility of broken symmetry states has also been explored in weakly disordered ABC--stacked trilayer graphene via a self--consistent Hartree--Fock approximation, with gapped broken symmetry states shown to be favored over both gapless and normal states\cite{PhysRevB.88.075408}. Gapped many--body states have also been investigated, of which we specifically mention quantum Hall states in chirally stacked systems\cite{PhysRevLett.106.156801,PhysRevB.80.165409}. 

This paper is structured as follows. 
In Sec. \ref{sec:TB_ham}, we begin by defining the tight binding model of the considered ABC--trilayer system and discuss its band structure. We then reduce the Hamiltonian to a nearest--neighbor only model as to simplify the Bethe--Salpeter calculations, discussing the dominant bands and the specific phase factors of each electronic state. 
Following from the single--particle regime, in Sec. \ref{sec:Bethe-Salpeter-Equation} we discuss the excitonic states of the system, obtained by solving the Bethe--Salpeter equation.
Finally, in Sec. \ref{sec:conductivity}, we discuss the optical response of the system. After outlining the method of computing the optical conductivity, we discuss the excitonic selection rules for both linearly polarized and circularly polarized light. We then consider additional hopping parameter in the Hamiltonian, discussing the resulting new selection rules and computing their contribution to the optical conductivity. Finally, we consider several different values of the external bias potential, computing the optical conductivity for each as to ascertain the tunability of the excitonic response. 

\section{Tight--Binding Model \label{sec:TB_ham}}

\begin{figure}[H]
	\centering
	\includegraphics[scale=0.35]{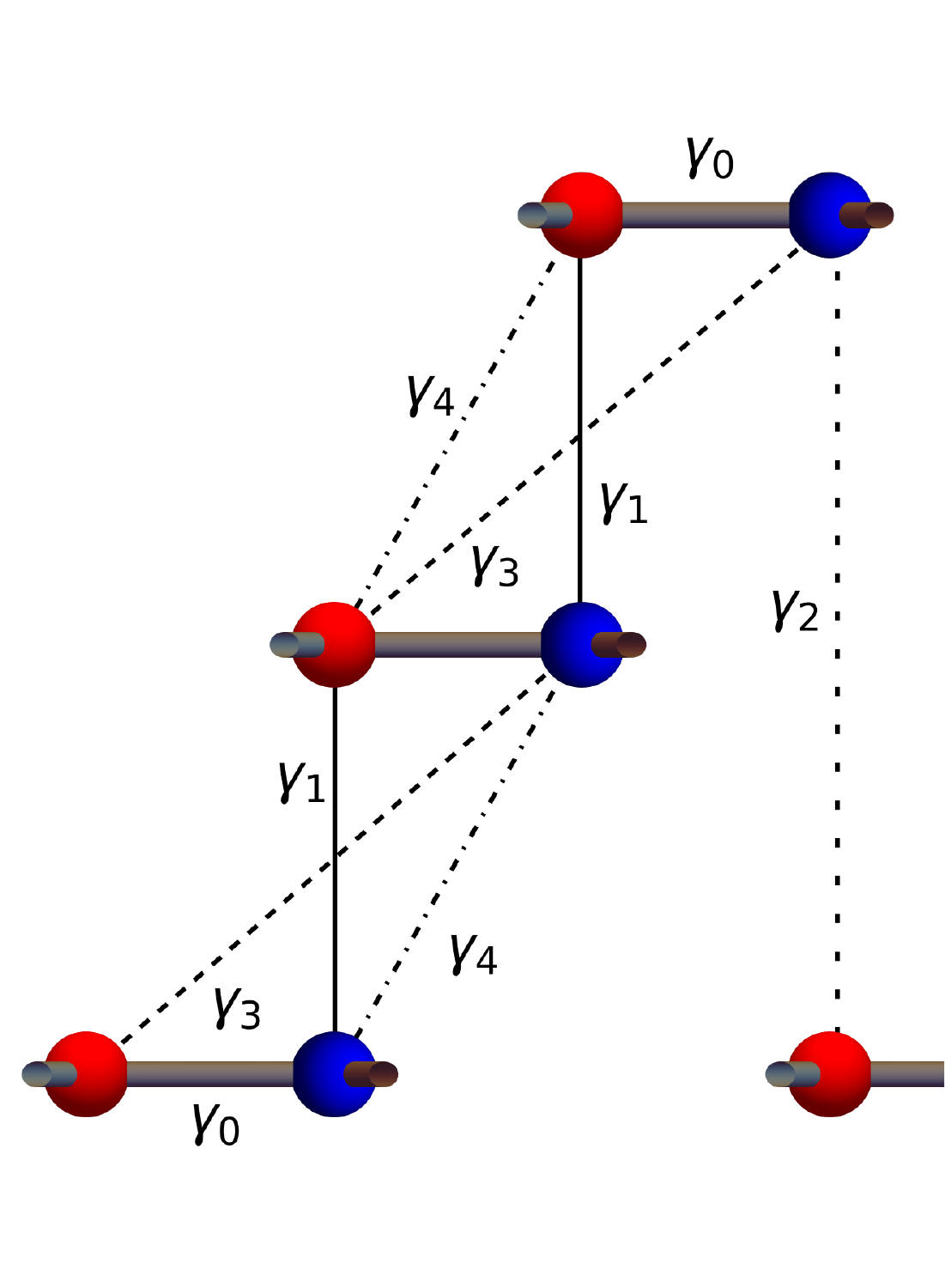}
	\caption{Schematic view of the hoppings included in the tight binding model. Red and blue dots represent the two different sublattices, while the different lines connecting them represent the different hopping terms considered. }\label{fig:lattice_stacking}
\end{figure}

For describing the excitonic properties of the ABC--stacked graphene trilayer, we first need to analyze the electronic properties of the system in the independent-electron approximation. This stacking order is characterized by the B sublattice of each layer laying opposite of the A sublattice of the layer above, but opposite to the honeycomb centers of the layer below, and is also known as rhombohedral stacking. We begin defining a tight binding Hamiltonian written directly in momentum space and taking into account the hoppings discussed in \cite{PhysRevB.88.075408}.  A schematic view of the hoppings considered is shown in Fig. \ref{fig:lattice_stacking}.

Throughout this paper, we will work in the $\left\{\ket{1,t},\ket{2,t},\ket{1,m},\ket{2,m},\ket{1,b},\ket{2,b}\right\} $ basis, where the $1/2$ labels represent the two sites in the monolayer graphene unit cell (red/blue dots in Fig. \ref{fig:lattice_stacking}, respectively) and the $t/m/b$ labels represent the top/middle/bottom layers. The tight binding Hamiltonian for ABC--stacked trilayer graphene for the hoppings shown in Fig. \ref{fig:lattice_stacking} can be written as \cite{PhysRevB.88.075408,PhysRevB.82.035409,doi:10.1021/nn3017926}
\begin{widetext}
\begin{equation}
	\mathcal{H}_{\mathrm{TB}}=\left[\begin{array}{cccccc}
		0 & \gamma_{0}\phi\left(\mathbf{k}\right) & \gamma_{4}\phi\left(\mathbf{k}\right) & \gamma_{3}\phi^{*}\left(\mathbf{k}\right) & 0 & \gamma_{2}\\
		\gamma_{0}\phi^{*}\left(\mathbf{k}\right) & 0 & \gamma_{1} & \gamma_{4}\phi\left(\mathbf{k}\right) & 0 & 0\\
		\gamma_{4}\phi^{*}\left(\mathbf{k}\right) & \gamma_{1} & 0 & \gamma_{0}\phi\left(\mathbf{k}\right) & \gamma_{4}\phi\left(\mathbf{k}\right) & \gamma_{3}\phi^{*}\left(\mathbf{k}\right)\\
		\gamma_{3}\phi\left(\mathbf{k}\right) & \gamma_{4}\phi^{*}\left(\mathbf{k}\right) & \gamma_{0}\phi^{*}\left(\mathbf{k}\right) & 0 & \gamma_{1} & \gamma_{4}\phi\left(\mathbf{k}\right)\\
		0 & 0 & \gamma_{4}\phi^{*}\left(\mathbf{k}\right) & \gamma_{1} & 0 & \gamma_{0}\phi\left(\mathbf{k}\right)\\
		\gamma_{2} & 0 & \gamma_{3}\phi\left(\mathbf{k}\right) & \gamma_{4}\phi^{*}\left(\mathbf{k}\right) & \gamma_{0}\phi^{*}\left(\mathbf{k}\right) & 0
	\end{array}\right],\label{eq:tb_model}
\end{equation}
\end{widetext}
with $\phi(\mathbf{k})$ obtained from the honeycomb geometry of the individual layers as 
\begin{equation}
	\phi(\mathbf{k})=e^{i k_{y} a / \sqrt{3}}\left[1+2 e^{-i 3 k_{y} a / 2 \sqrt{3}} \cos \left(\frac{k_{x} a}{2}\right)\right]
\end{equation}
and $a=2.46\,\text{\AA}$ the carbon--carbon distance in graphene.

As we are interested in the low energy response of the system, we restrict our study to the Dirac points of the first Brillouin zone. Close to these Dirac points, $\phi\left(\mathbf{k}\right)$ can be approximated as 
\[
\phi\left(\mathbf{k}\right)\approx\frac{3}{2}a\,\tau ke^{i\tau\theta},
\]
with $ \tau=\pm1$ the Dirac valley index, $k=\left|\mathbf{k}\right|$ and $\theta=\arctan\left(\frac{k_{y}}{k_{x}}\right)$.

\begin{figure}
	\begin{centering}
		\includegraphics[scale=0.66]{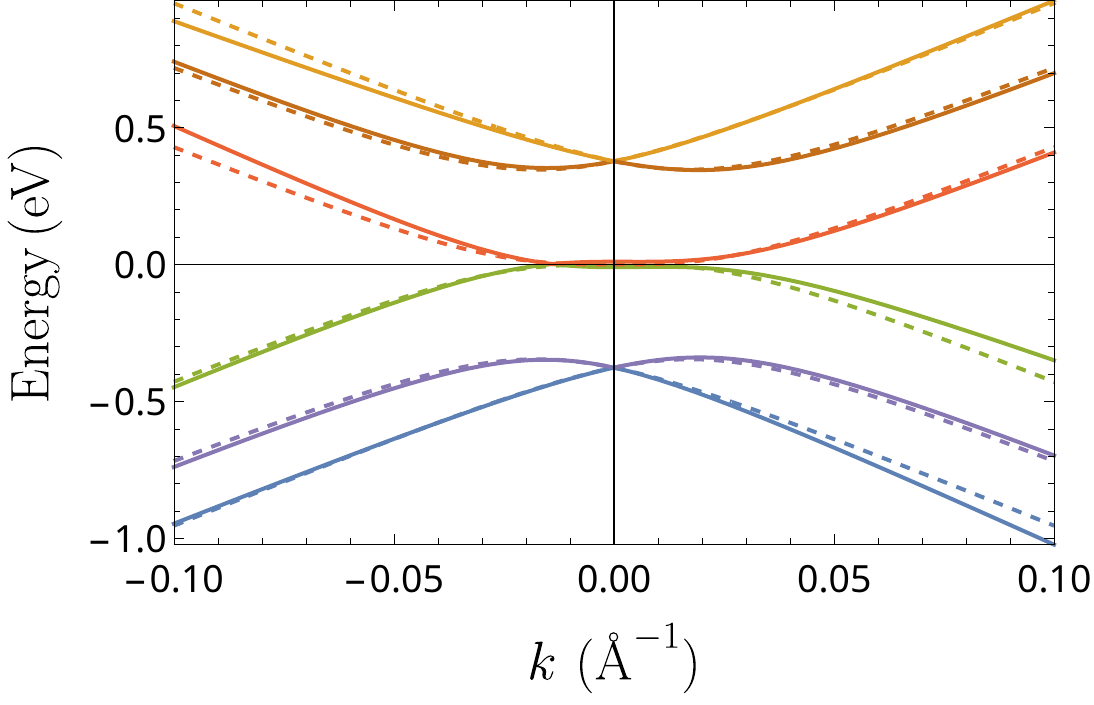}
		\par\end{centering}
	\caption{Electronic bands near the Dirac valley $\tau=1$ for ABC--stacked trilayer
		graphene. Solid lines represent the full tight binding Hamiltonian of Eq. (\ref{eq:tb_model}), while dashed lines represent the minimal model Hamiltonian where only the hopping parameters $ \gamma_{0} $ and $ \gamma_{1} $ were considered. }\label{fig:band_struct_ABC_full}
\end{figure}

The nearest-neighbor intralayer and interlayer hopping processes $\gamma_{0}$
and $\gamma_{1}$ are responsible for the general features of the band structure,
while $\gamma_{2}$, $\gamma_{4}$ and the trigonal warping $\gamma_{3}$ parameter have their main impact
close to the band-crossing points. Considering the graphite hopping parameter values described in \cite{PhysRevB.88.075408}, given by $\gamma_{0}=3.12\,\mathrm{eV}$, $\gamma_{1}=0.377\,\mathrm{eV}$, $\gamma_{2}=0.01\,\mathrm{eV}$ and  $\gamma_{3}=0.3\,\mathrm{eV}$, as well as the $ \gamma_{4} $ hopping parameter described in \cite{PhysRevB.82.035409}, $\gamma_{4}=-0.1\,\mathrm{eV}$, the band structure near the one of the two Dirac points is given in Fig. \ref{fig:band_struct_ABC_full}. 
In this figure, the band structure for a minimal model Hamiltonian where $ \gamma_{2}=\gamma_{3}=\gamma_{4}=0 $ is also plotted in dashed lines. The agreement between the full and the minimal models is quite good. No bandgap is present in either model, with the two lowest energy bands intersecting at $ k\approx-0.014\,\text{\AA}^{-1} $ in the full model and at $ k=0 $ in the minimal model. This intersection of the lowest energy bands in the full model is similar to that which is present at $ k=0 $ for the two higher energy bands, with no bandcrossing occurring. Focusing on the higher energy bands, their previously mentioned intersection at $ k=0 $ occurs at an energy of roughly $ 380\,\mathrm{meV} $ (see Fig. \ref{fig:band_struct_ABC_full}, brown/orange lines for conduction bands and purple/blue lines for valence bands). The minimum of these two higher energy bands occurs at $ 350\,\mathrm{meV} $, significantly higher than the energy scale of the lowest energy bands. 

\subsection{Nearest--Neighbor Biased Hamiltonian}

Since there are no significant differences between the full tight binding Hamiltonian and the minimal model close to $ k=0 $, we consider, for matters of simplicity, only the minimal model with $ \gamma_0 $ and $ \gamma_1 $ both finite. The adoption of this minimal model for the electronic motion in the ABC--trilayer graphene allows, as discussed ahead, separation of variables in the eigenvectors of the tight binding Hamiltonian (see Eq. (\ref{eq:generic_ev})), greatly simplifying the momentum integration in the Bethe--Salpeter equation.  The effects of considering non--zero trigonal warping on the optical selection rules, \emph{i.e.} setting $\gamma_{3}=0.3\,\mathrm{eV}$, will be discussed in Sec. \ref{sec:trigonal_warp}. 

\begin{figure}
	\begin{centering}
		\includegraphics[scale=0.66]{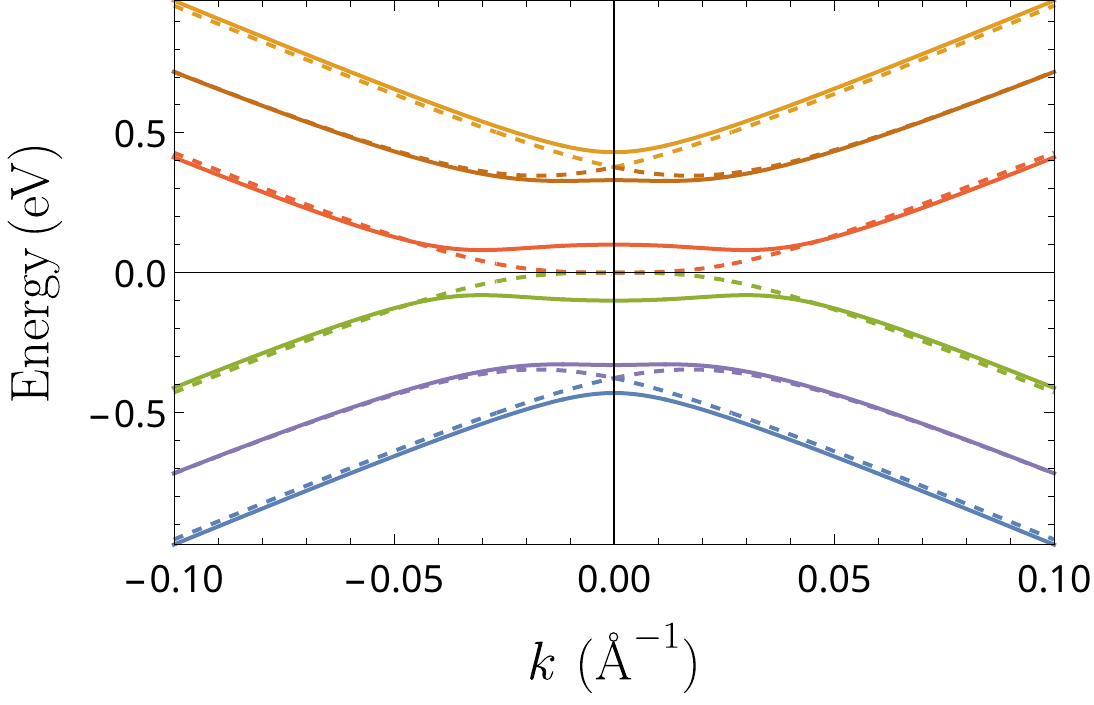}
	\par\end{centering}
	\caption{Electronic bands near the Dirac valley $\tau=1$ for a minimal model of biased ABC--stacked trilayer graphene with bias potential $V=0\,\mathrm{meV}$ (dashed lines) and $V=100\,\mathrm{meV}$ (solid lines).}\label{fig:band_struct_ABC}
\end{figure}

Adding an external electric field perpendicular to the layers introduces in the Hamiltonian an additional term, which takes into account the electric potential in the different layers. The new Hamiltonian reads 
\begin{equation}
	\mathcal{H}=\mathcal{H}_{\mathrm{TB}}+V_{\mathrm{diag}}\left[1,1,0,0,-1,-1\right],\label{eq:biased_ham}
\end{equation}
where $ V_{\mathrm{diag}}\left[1,1,0,0,-1,-1\right] $ represents a diagonal matrix where the diagonal elements are those in square brackets (\emph{i.e.}, $ \left[1,1,0,0,-1,-1\right] $), the rest of the elements being zero. This corresponds to an electric potential of $+V$ in the top layer, $0$ in the middle layer, and $-V$ in the bottom layer, meaning that the total potential difference between the top and bottom layers will be $ 2V $. The band structure near the Dirac point of the $ \tau=1 $ valley for $V=0$ and
$100\,\mathrm{meV}$ is given in Fig. \ref{fig:band_struct_ABC}

As expected\cite{Lui2011}, a gap of $E_{gap}=2V$ opens at the Dirac point, although that is not the smallest gap in the system. For the bias potential considered in Fig. \ref{fig:band_struct_ABC} ($ 100\,\mathrm{meV} $), a gap of $\Delta=160\,\mathrm{meV}$  exists at roughly $k\approx \pm0.03\, \text{\AA} ^{-1}$. This second, smaller gap remains the smallest for all finite values of the bias potential, although its location depends on the value of $V$ (minimum at $\pm0.0085\, \text{\AA} ^{-1}$ for $V=10\,\mathrm{meV}$, and at $\pm0.05\, \text{\AA} ^{-1}$ for $V=250\,\mathrm{meV}$). Additionally, when a bias potential is introduced in the system, a gap also appears between the two higher energy bands, removing the intersection at $ k=0 $ visible in the dashed lines.

As the characteristic polynomial of this Hamiltonian is of order six, the exact form of the eigenvector for each of the six bands is cumbersome.
As such, we will not write their explicit expressions. Instead, and as they have a well--defined phase
in each of the six spinor components of each eigenvector, we will extract this phase factor explicitly. This separation will prove useful for solving the Bethe--Salpeter equation, allowing us to transform the two dimensional integral into a 1D problem.
This generic eigenvector will then be given by 
\begin{widetext}
\begin{align}
	\left|u_{\mathbf{k}}^{v,\eta}\right\rangle&=\left[e^{3i\theta\tau}\psi_{1,v}^{\eta},e^{2i\theta\tau}\psi_{2,v}^{\eta},e^{2i\theta\tau}\psi_{3,v}^{\eta},e^{i\theta\tau}\psi_{4,v}^{\eta},e^{i\theta\tau}\psi_{5,v}^{\eta},\psi_{6,v}^{\eta}\right]^{\intercal}\nonumber\\
	\left|u_{\mathbf{k}}^{c,\eta}\right\rangle&=\left[e^{3i\theta\tau}\psi_{1,c}^{\eta},e^{2i\theta\tau}\psi_{2,c}^{\eta},e^{2i\theta\tau}\psi_{3,c}^{\eta},e^{i\theta\tau}\psi_{4,c}^{\eta},e^{i\theta\tau}\psi_{5,c}^{\eta},\psi_{6,c}^{\eta}\right]^{\intercal},\label{eq:generic_ev}
\end{align}
\end{widetext}
where the $ k $ dependence has been included in the radial $ \psi_{j,c/v}^{\eta} $ radial spinor components for compactness, $ c/v $ distinguishes between conduction and valence bands, and $ \eta $ is the band index that distinguishes the three individual bands in each set ($ \eta=-1 $ for the band closest to the gap, $ \eta=0 $ for the intermediate band and $ \eta=+1 $ for the band furthest from the gap, see Fig. \ref{fig:band_struct_ABC}).

However, due to the definition of the angular variable $ \theta $, the complex exponential $ e^{i\theta} $ becomes ill--defined and discontinuous as $ k\rightarrow 0$. To avoid this discontinuity, we group the phase factors such that complex exponentials only appear multiplied by terms that vanish at $ k=0 $, removing numerical difficulties stemming from this discontinuity\cite{PhysRevB.105.045411}. This will lead to different forms of the eigenvectors from Eq. (\ref{eq:generic_ev}) depending on the specific band, given in generic fashion in Eq. (\ref{eq:eigenvectors_fix}).
\begin{widetext}
\begin{align}
	&\left|u_{\mathbf{k}}^{c,-1}\right\rangle=\left[\psi_{1,c}^{-},e^{-i\theta\tau}\psi_{2,c}^{-},e^{-i\theta\tau}\psi_{3,c}^{-},e^{-2i\theta\tau}\psi_{4,c}^{-},e^{-2i\theta\tau}\psi_{5,c}^{-},e^{-3i\theta\tau}\psi_{6,c}^{-}\right]^{\intercal}\nonumber\\
	&\left|u_{\mathbf{k}}^{v,-1}\right\rangle=\left[e^{3i\theta\tau}\psi_{1,v}^{-},e^{2i\theta\tau}\psi_{2,v}^{-},e^{2i\theta\tau}\psi_{3,v}^{-},e^{i\theta\tau}\psi_{4,v}^{-},e^{i\theta\tau}\psi_{5,v}^{-},\psi_{6,v}^{-}\right]^{\intercal}\nonumber\\
	&\left|u_{\mathbf{k}}^{c,0}\right\rangle=\left[e^{2i\theta\tau}\psi_{1,c}^{0},e^{i\theta\tau}\psi_{2,c}^{0},e^{i\theta\tau}\psi_{3,c}^{0},\psi_{4,c}^{0},\psi_{5,c}^{0},e^{-i\theta\tau}\psi_{6,c}^{0}\right]^{\intercal}\nonumber\\
	&\left|u_{\mathbf{k}}^{v,0}\right\rangle=\left[e^{i\theta\tau}\psi_{1,v}^{0},\psi_{2,v}^{0},\psi_{3,v}^{0},e^{-i\theta\tau}\psi_{4,v}^{0},e^{-i\theta\tau}\psi_{5,v}^{0},e^{-2i\theta\tau}\psi_{6,v}^{0}\right]^{\intercal}\nonumber\\
	&\left|u_{\mathbf{k}}^{c,+1}\right\rangle=\left[e^{i\theta\tau}\psi_{1,c}^{+},\psi_{2,c}^{+},\psi_{3,c}^{+},e^{-i\theta\tau}\psi_{4,c}^{+},e^{-i\theta\tau}\psi_{5,c}^{+},e^{-2i\theta\tau}\psi_{6,c}^{+}\right]^{\intercal}\nonumber\\
	&\left|u_{\mathbf{k}}^{v,+1}\right\rangle=\left[e^{2i\theta\tau}\psi_{1,v}^{+},e^{i\theta\tau}\psi_{2,v}^{+},e^{i\theta\tau}\psi_{3,v}^{+},\psi_{4,v}^{+},\psi_{5,v}^{+},e^{-i\theta\tau}\psi_{6,v}^{+}\right]^{\intercal}.\label{eq:eigenvectors_fix}
\end{align}
\end{widetext}
This phase choice of the Bloch factors will play a crucial role in determining the optical selection rules and leads to Hydrogen--like selection rules in the monolayer\cite{PhysRevB.105.045411}. It is important to note, however, that this choice of phase factors breaks down for sufficiently large values of the bias potential. At $ V\approx260\,\mathrm{meV} $ the phases of the $ \eta=-1 $ and $ \eta=0 $ begin mixing as the top of the $ \eta=-1 $ band becomes extremely close to the bottom of the $ \eta=0 $ band.
As such, we end our calculations at $ V=110\,\mathrm{meV} $ as to be sufficiently far away from this regime.

Trigonal warping was not included in the minimal model Hamiltonian as its presence makes separating the phase factor of each spinor entry similarly to Eqs. (\ref{eq:generic_ev}--\ref{eq:eigenvectors_fix}) impossible. Ignoring trigonal warping at this level is not a stringent approximation as shows the results of Fig. \ref{fig:band_struct_ABC_full}. Still, we will consider its contribution to the dipole moment operator when selection rules are discussed as it leads to important new optical selection rules. 

\section{Bethe--Salpeter Equation \label{sec:Bethe-Salpeter-Equation}}

Having finalized the discussion of the electronic band structure, we will now move on to the excitonic states. To compute the excitonic wave functions and their binding energies we will solve the Bethe--Salpeter equation. For a multi--band system, the Bethe--Salpeter equation can be written
in momentum space as\cite{PhysRevB.99.235433,PhysRevB.92.235432,PhysRevLett.120.087402,PhysRevB.104.115120}
\begin{widetext}
\begin{align}
E \, \psi_{c,\eta_{1};v,\eta_{4}}\left(\mathbf{k}\right) & =\left(E_{\mathbf{k}}^{c,\eta_{1}}-E_{\mathbf{k}}^{v,\eta_{4}}\right)\psi_{c,\eta_{1};v,\eta_{4}}\left(\mathbf{k}\right)+\label{eq:BSE-simplified}\\
&\quad+\sum_{\eta_{2},\eta_{3}}\sum_{\mathbf{q}}V\left(\mathbf{k}-\mathbf{q}\right)\left\langle u_{\mathbf{k}}^{c,\eta_{1}}\mid u_{\mathbf{q}}^{c,\eta_{2}}\right\rangle \left\langle u_{\mathbf{q}}^{v,\eta_{3}}\mid u_{\mathbf{k}}^{v,\eta_{4}}\right\rangle\psi_{c,\eta_{2};v,\eta_{3}}\left(\mathbf{q}\right) \nonumber
\end{align}
\end{widetext}
where
$ \psi_{c,\eta_{1};v,\eta_{4}}\left(\mathbf{k}\right) $ is the excitonic wave function that we wish to obtain,  $\left|u_{\mathbf{k}}^{v/c,\eta}\right\rangle $ and $E_{\mathbf{k}}^{v/c,\eta}$
are the single particle electronic wave functions (Eqs. (\ref{eq:generic_ev}--\ref{eq:eigenvectors_fix})) and energies, respectively, and $V\left(\mathbf{k}\right)$
is an electrostatic potential coupling different bands and thus capturing
many--body effects including the intrinsic many--body nature of excitons. 

In this paper, we consider the electrostatic potential to be the Rytova--Keldysh
potential \cite{rytova1967,keldysh1979coulomb} (usually employed to describe excitonic phenomena in mono- and few--layer
materials), which can be obtained by solving the Poisson equation for a charge embedded in a thin film of vanishing thickness. In momentum space, this potential is given by 
\[
V\left(\mathbf{k}\right)=\frac{\hbar c\alpha}{\epsilon}\frac{1}{k\left(1+r_{0}k\right)},
\]
where $\alpha=1/137$ is the fine--structure constant, $\epsilon$
the mean dielectric constant of the medium above/below the ABC--trilayer graphene. The parameter $r_{0}$ corresponds to an in--plane screening length related
to the 2D polarizability of the material. It can be calculated from the single particle Hamiltonian of the system, although \emph{ab initio} calculations might be necessary for accurate computation of $ r_0 $ depending on the material\cite{acs.nanolett.9b02982}. This screening parameter varies with the bias potential $ V $, and its numerical value is of the utmost importance if the excitonic properties of a specific system are to be studied accurately\cite{PhysRevB.92.245123,sponza2020proper}. An in--depth discussion of the in--plane screening length in bilayer graphene has been done in Ref. \cite{PhysRevB.99.035429}, and we perform a simplified version of this procedure for ABC--trilayer graphene in Appendix \ref{app:Effective-Screening-Length}. 

To solve the Bethe--Salpeter equation, we assume that the excitons
have a well--defined angular momentum quantum number $m$, such that
their wave functions can be written as $\psi_{c,\eta_{1};v,\eta_{4}}\left(\mathbf{k}\right)=f_{c,\eta_{1};v,\eta_{4}}\left(k\right)e^{im\theta}$. 
Furthermore, it is important to note that Eq. (\ref{eq:BSE-simplified}) is actually a separate equation for each pair of bands $ c,\eta_{1};v,\eta_{4} $. This implies that there are $ 9 $ equations (3 valence times 3 conduction) that must be solved, stemming from the three valence and three conduction bands.  
Additionally, as mentioned previously, a careful choice of the phases of the single--particle spinors allows us to transform the BSE into a $ 1D $ integral equation. Both the discussion on the necessary transformations to solving the Bethe--Salpeter equation in biased ABC--trilayer graphene and the description of the numeric methodology are available in Appendix \ref{app:BSE}. Solving this eigenvalue problem, one obtains the excitonic eigenvalues and eigenfunctions. 

Having determined the solutions for a wide range of biases, we observed that of the 9 sets of $  \psi_{c,\eta_{1};v,\eta_{4}}\left(\mathbf{k}\right) $, those corresponding to $ \eta_{1} = \eta_{4} = -1$ were by far the dominant contributions. This is a reasonable and somewhat expected result, as intuition tells us that the bands close to the gap should dominate the system's low--energy response. As such, calculations can be greatly optimized by restricting the sum over bands to only the $ \eta=-1 $ bands. It is important to note that, as the bias potential increases past a certain point (roughly $ V\approx200\,\mathrm{meV}$), the $ \eta=-1 $ bands are no longer the sole dominant contribution. At this external bias, one must also take into account the next pair of bands to obtain a reasonable result, greatly increasing the computational complexity and calculation time.

When discussing excitonic states, we adopt a nomenclature similar to what is used in the Hydrogen atom, with states with angular momentum $ m=0 $ being $ s $--series states, states with angular momentum $ \left|m\right|=1 $ being $ p $--series states, and analogously to higher angular momenta. To distinguish $\pm m$ states, for $ m\neq 0 $, we will use the sign of the angular momentum in index (\emph{i.e.}, $ 3d_+ $ and $ 3d_- $ states).

\begin{figure*}
	\centering
	\begin{minipage}[c]{0.3\linewidth}
		\includegraphics[scale=0.8]{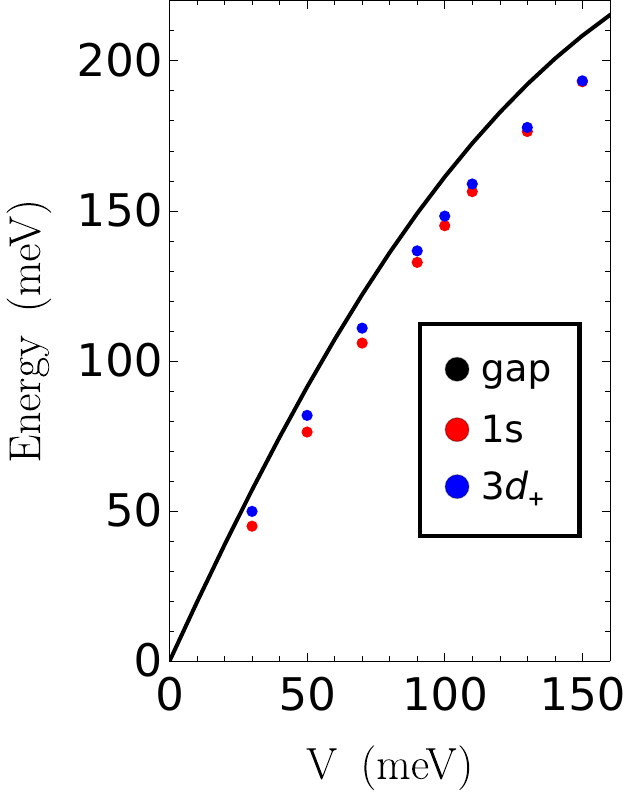}
	\end{minipage}
	\quad\;
	\begin{minipage}[c]{0.655\linewidth}
		\includegraphics[scale=0.35]{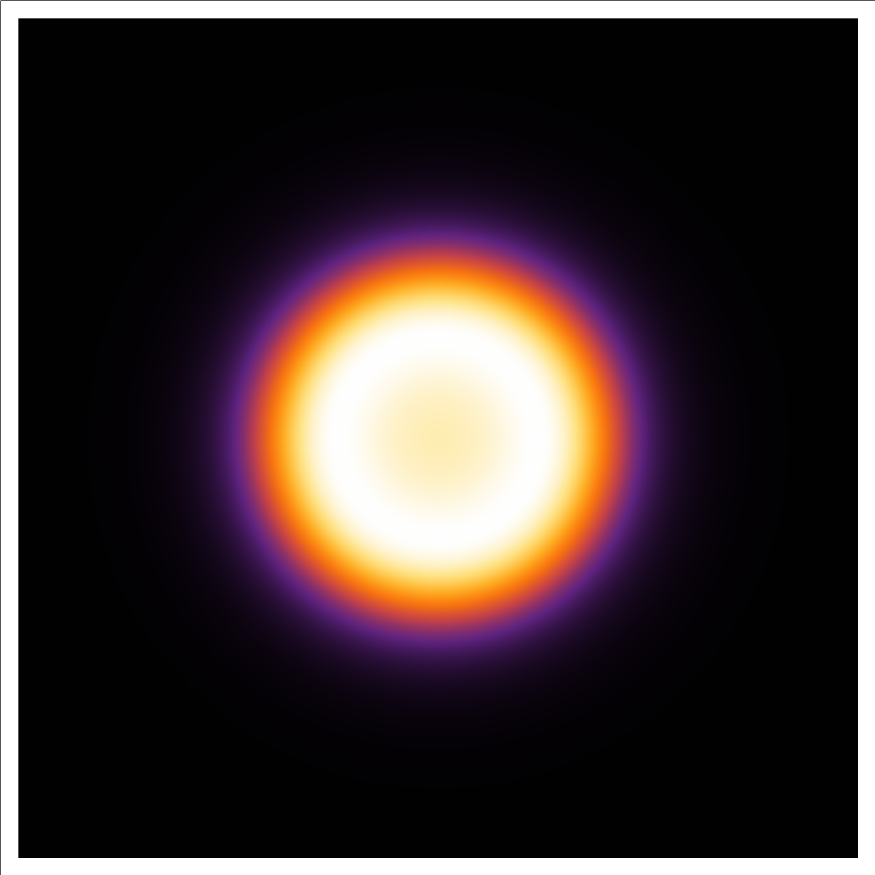}
		\includegraphics[scale=0.35]{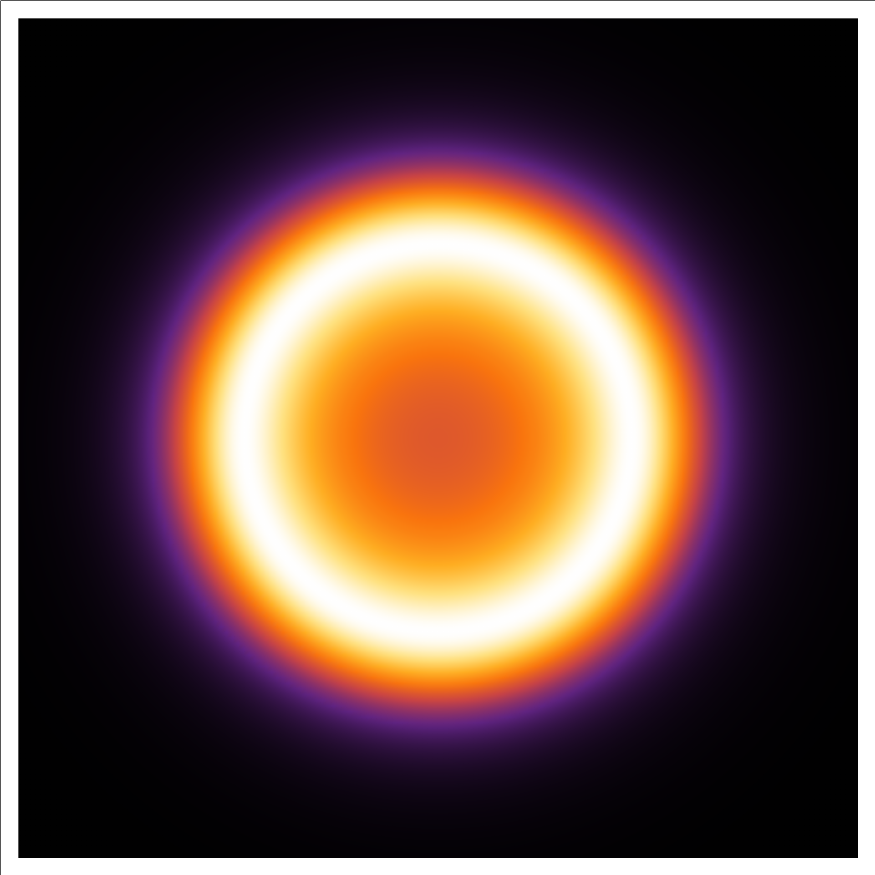}
		\includegraphics[scale=0.35]{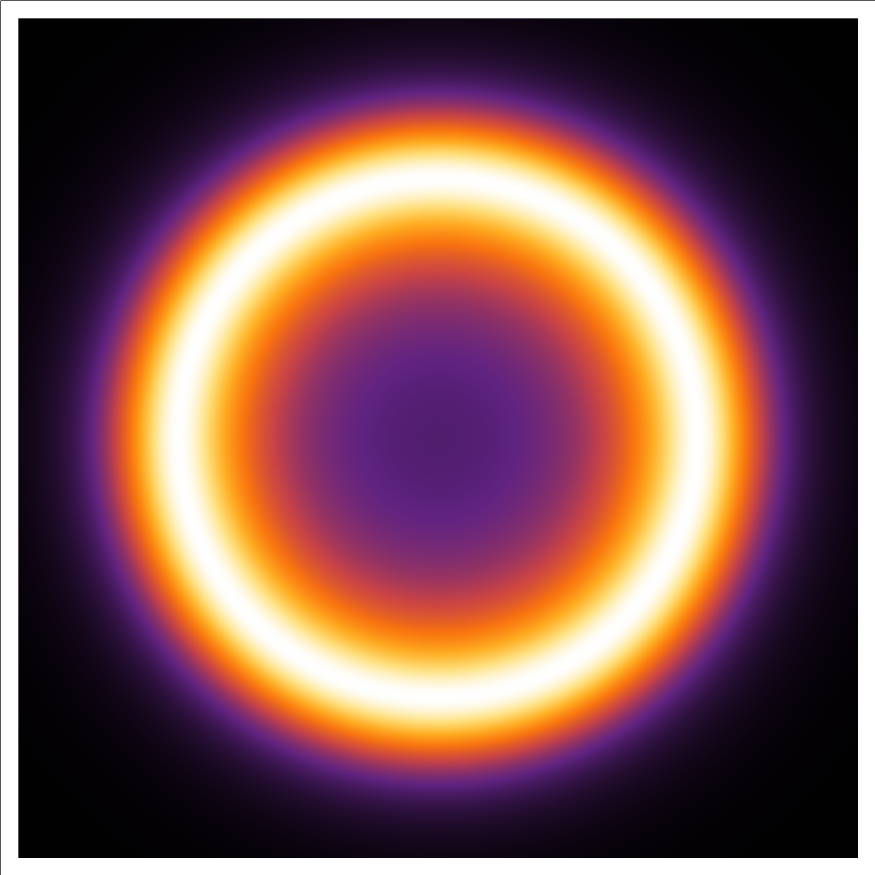}
		
		\includegraphics[scale=0.35]{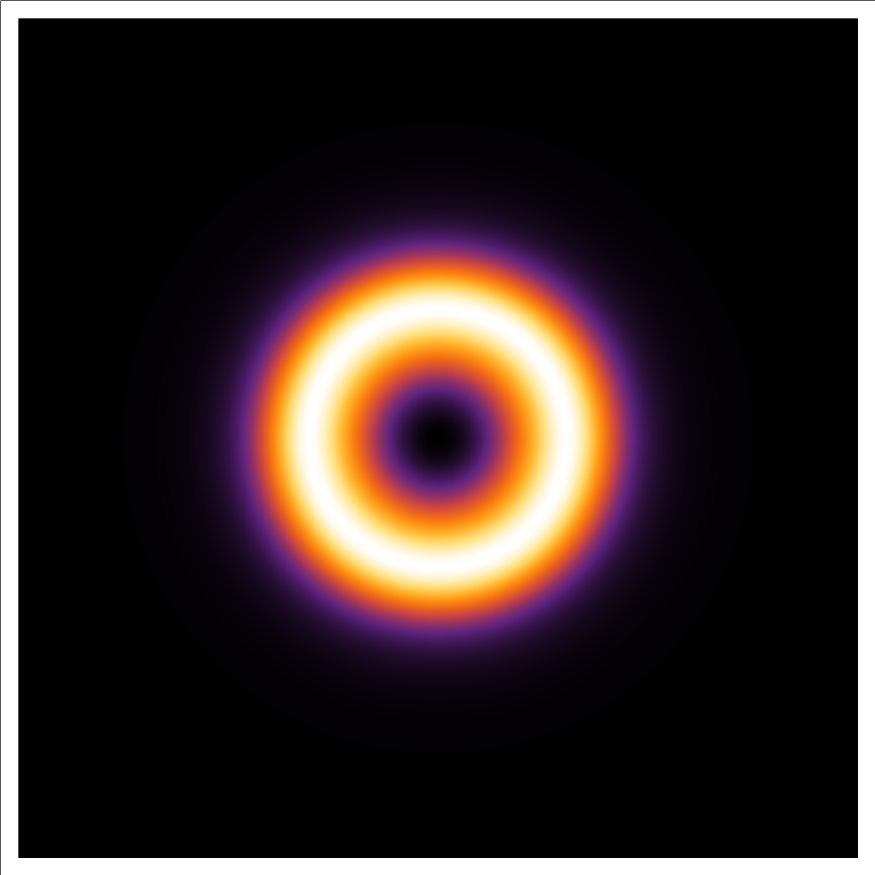}
		\includegraphics[scale=0.35]{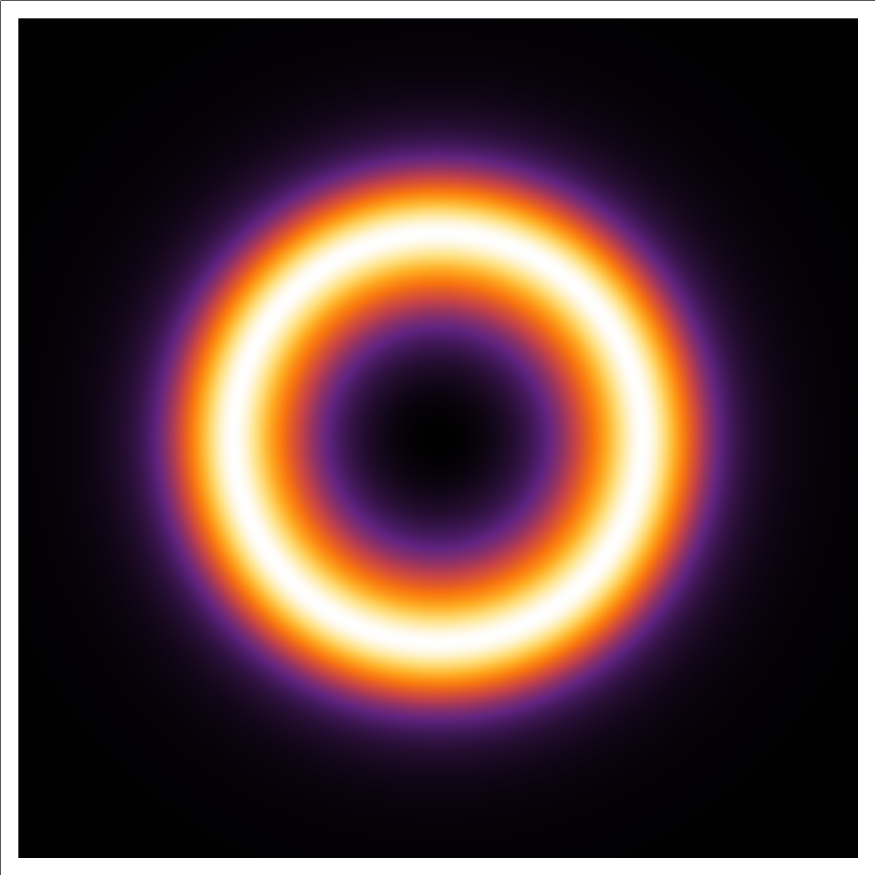}
		\includegraphics[scale=0.35]{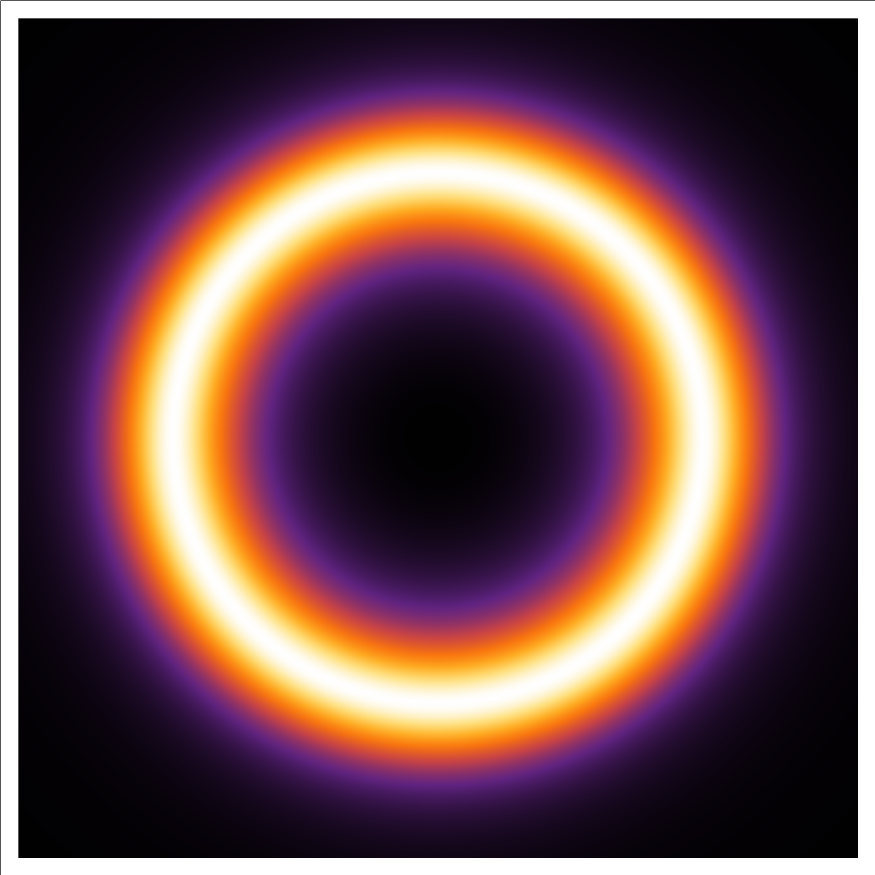}
	\end{minipage}
	\caption{Left: Electronic bandgap (black line), together with binding energies of the (optically dark) excitonic $ 1s $ ground state (red dots) and (optically bright) excitonic $ 3d_+ $ state, for various bias potentials between $V=30\,\mathrm{meV}$  and $V=150\,\mathrm{meV}$. Right: Density plot of the absolute value squared of the excitonic $ 1s $ (top panels) and $ 3d_{+} $ (bottom panels) wave functions in ABC--stacked trilayer graphene encapsulated in hBN with various external biases $V=30\,\mathrm{meV}$, $V=70\,\mathrm{meV}$ and $V=110\,\mathrm{meV}$. The region plotted in each panel is a square of side $0.1\,\text{\AA}^{-1}$ centered at $ k=0 $.}\label{fig:density_plot_V_multi}
\end{figure*}

To finalize this section, we depict the density plot of the $ 1s $ and the $ 3d_+ $ excitonic states for three different external bias in Fig. \ref{fig:density_plot_V_multi}, together with the binding energies of the two excitonic states in question and the electronic bandgap. The $ 1s $ state is presented only for comparison, as it is the only state that is non--zero at $ k=0 $ and, as such, is sufficiently distinct from all other excitonic states. However, it is optically dark and will play no part in the optical conductivity, as we will show in Sec. \ref{sec:conductivity}.

As the external field $ V $ increases, the effective screening length $ r_0 $ decreases (see Appendix \ref{app:Effective-Screening-Length}) leading to more tightly bound excitons in real space. This is clear by the increase in the delocalization in momentum space of the excitonic wave functions as the field increases, which is equivalent to localization in real space, in accordance with what was expected from the decrease in screening length. Considering the wave function for $ V=110\,\mathrm{meV} $, the spread in momentum space of the $ 3d_{+} $ states is roughly $ \Delta k \approx 0.08\,\text{\AA}^{-1}$. As such, the spread in real space will be approximately $ \Delta r = 2\pi/\Delta k\approx80\,\text{\AA} $.

Having obtained the excitonic wave functions and discussed their dependence on the external bias potential, we will now compute the optical linear conductivity and discuss the selection rules obtained from the tight binding Hamiltonian.

\section{Excitonic Conductivity \label{sec:conductivity}}

In this section, our goal is to obtain the excitonic linear optical conductivity for biased trilayer graphene, followed by discussing the tunability of the obtained resonances via the external potential. We will begin by determining the optical selection rules of our system and the impact of each hopping term in the Hamiltonian on these same selection rules. This discussion will be focused on both linearly polarized light and circularly polarized light, while the final computations will focus only on linearly polarized light, as circular polarization does not generate new possible transitions. 

In the dipole approximation, and considering normal incidence, the optical conductivity
is given by\cite{PhysRevB.92.235432} 
\begin{equation}
\sigma_{\alpha,\beta}\left(\hbar\omega\right)\propto \sum_{n} \frac{\boldsymbol{\Omega}_{n,\alpha} \boldsymbol{\Omega}_{n,\beta}^{*}}{E_{n}-\hbar \omega-i\Gamma_n}+(\omega \rightarrow-\omega)^{*},\label{eq:opt_cond}
\end{equation}
where the sum over $n$ represents the sum over excitonic states with energy $ E_n $ and wave function $ \psi_{n,cv} $, and $ \Gamma_n $ is a phenomenological broadening parameter considered to be $ n $--dependent in a similar fashion as \cite{PhysRevB.105.045411}.
In Eq. (\ref{eq:opt_cond}), $\boldsymbol{\Omega}_{n,\alpha}$ is defined as
\begin{equation}
	\boldsymbol{\Omega}_{n,\alpha}=\sum_{c,v}\sum_{\mathbf{k}}\psi_{n,cv}\left(\mathbf{k}\right)\left\langle u_{\mathbf{k}}^{v}\left|\mathbf{r}_{\alpha}\right|u_{\mathbf{k}}^{c}\right\rangle ,\label{eq:tg_pedersen_OMEGA}
\end{equation}
with $\left\langle u_{\mathbf{k}}^{v}\left|\mathbf{r}_{\alpha}\right|u_{\mathbf{k}}^{c}\right\rangle$
the interband dipole operator matrix element in the $ \alpha $ direction, obtained using the relation
\[
\left\langle u_{\mathbf{k}}^{v}\left|\mathbf{r}_{\alpha}\right|u_{\mathbf{k}}^{c}\right\rangle =\frac{\left\langle u_{\mathbf{k}}^{v}\left|\left[H,\mathbf{r}_{\alpha}\right]\right|u_{\mathbf{k}}^{c}\right\rangle }{E_{k}^{v}-E_{k}^{c}}.
\]
Knowing this relation, one then expands the commutator $ \left\langle u_{\mathbf{k}}^{v}\left|\left[H,\mathbf{r}_{\alpha}\right]\right|u_{\mathbf{k}}^{c}\right\rangle $ and the optical selection rules are directly obtained from the phase factors of the single particle states in Eq. (\ref{eq:eigenvectors_fix}). With these phase factors fixed, one can then study what optical transitions become allowed when specific hopping terms are included in the tight binding Hamiltonian.  

As mentioned previously, although only the nearest--neighbor hopping terms were considered when solving the Bethe--Salpeter equation, the effects of the trigonal warping hopping $ \gamma_{3} $ will also be taken into account during the evaluation of the commutator $\left[H,\mathbf{r}_{\alpha}\right]$ as it plays a crucial role in the system's optical selection rules. The impact of the $ \gamma_{4} $ hopping parameter will also be discussed, even though it does not generate new selection rules. The magnitude of this hopping parameter is also much smaller than that of $ \gamma_{0} $, leading to no significant change in the excitonic peaks. As such, we will not include its contribution in the final optical conductivity.

\subsection{Linearly and Circularly Polarized Light\label{sec:lin_pol}}

Considering linearly polarized light, fixed (without loss of generality) in the $x$ direction, and taking again the thermodynamic limit, we write Eq. (\ref{eq:tg_pedersen_OMEGA}) as
\begin{equation}
	\boldsymbol{\Omega}_{n}=\sum_{c,v}\int\psi_{n,cv}\left(\mathbf{k}\right)\frac{\left\langle u_{\mathbf{k}}^{v}\left|\left[H,x\right]\right|u_{\mathbf{k}}^{c}\right\rangle }{E_{k}^{v}-E_{k}^{c}}k\,dk\,d\theta.\label{eq:omega_integral}
\end{equation}
As such, the optical conductivity will be given by
\begin{widetext} 
\begin{equation}
	\sigma_{xx}\left(\hbar\omega\right)=\frac{i}{2 \pi^{3}}\sum_{n}\left[\frac{1}{E_{n}-\left(\hbar\omega+i\Gamma_n\right)}\left|\sum_{c,v}\int\psi_{n,cv}\left(\mathbf{k}\right)\frac{\left\langle u_{\mathbf{k}}^{v}\left|\left[H,x\right]\right|u_{\mathbf{k}}^{c}\right\rangle }{E_{k}^{v}-E_{k}^{c}}k\,dk\,d\theta\right|^{2}\right]+\left(\omega\rightarrow-\omega\right)^{*},
\end{equation}
\end{widetext}
which can be quickly computed as solving the Bethe--Salpeter equation
provided us with both $E_{n}$ and $\psi_{n,cv}\left(\mathbf{k}\right)$, and the diagonalization of Eq. (\ref{eq:biased_ham}) provides us with the band structure. 

As was discussed in Sec. \ref{sec:Bethe-Salpeter-Equation}, we can safely discard the contribution from higher energy bands and focus only on the two electronic bands closest to the gap as long as the external bias remains sufficiently small. This somewhat simplifies the previous expression, with the optical conductivity being given by
\begin{widetext} 
\begin{equation}
	\sigma_{xx}\left(\hbar\omega\right)=\frac{i}{2 \pi^{3}}\sum_{n}\left[\frac{1}{E_{n}-\left(\hbar\omega+i\Gamma_n\right)}\left|\int\psi_{n}\left(\mathbf{k}\right)\frac{\left\langle u_{\mathbf{k}}^{v}\left|\left[H,x\right]\right|u_{\mathbf{k}}^{c}\right\rangle }{E_{k}^{v}-E_{k}^{c}}k\,dk\,d\theta\right|^{2}\right]+\left(\omega\rightarrow-\omega\right)^{*},
\end{equation}
\end{widetext}
where $ \psi_{n} $ corresponds to the excitonic wave functions when only these two lowest energy bands are considered. 

The optical selection rules are now evident, as the integral of Eq. (\ref{eq:omega_integral}) is only non--zero for states with angular momentum symmetric to the phase factors obtained by expanding the commutator $ \left\langle u_{\mathbf{k}}^{v}\left|\left[H,x\right]\right|u_{\mathbf{k}}^{c}\right\rangle $. Explicitly expanding this commutator, we obtain
\begin{align}
	\left\langle u_{\mathbf{k}}^{v}\left|\left[H,x\right]\right|u_{\mathbf{k}}^{c}\right\rangle & \propto\mathcal{A}^{\gamma_{0}}_{c,v}e^{-2i \tau \theta}+\mathcal{A}^{\gamma_{0}}_{v,c}e^{-4i \tau \theta},
\end{align} 
with
\begin{align*}
	\mathcal{A}^{\gamma_{0}}_{c,v}&=\psi_{1,c}\psi_{2,v}+\psi_{3,c}\psi_{4,v}+\psi_{5,c}\psi_{6,v},\\
	\mathcal{A}^{\gamma_{0}}_{v,c}&=\psi_{1,v}\psi_{2,c}+\psi_{3,v}\psi_{4,c}+\psi_{5,v}\psi_{6,c}.
\end{align*}
In this expression, we can see that the first term only leads to a non--zero contribution for $m=2\tau$  states ($ d $ series), and the second term for $m=4\tau$ states ($ g $ series). Focusing on the $ \tau=1 $ valley and comparing the relative amplitudes of the contributions from both series, those from $d_{+}$--series states dominate and $g_{+}$--series states go totally unnoticed, with the relative amplitude being less than $ 0.1\% $ for an external bias of $ V=30\,\mathrm{meV} $.

For circularly polarized light, the procedure is equivalent to that which was performed above, with the only slight change being the different interband dipole operator matrix element. In this regime, this operator will be written as 
$\left\langle u_{\mathbf{k}}^{v}\left|\left[H,x\pm iy\right]\right|u_{\mathbf{k}}^{c}\right\rangle$, 
with $\pm$ differentiating between right polarization $\left(+\right) $ and left polarization $\left(-\right)$. 
Focusing on right polarized light for simplicity, the full interband dipole operator can be written as 
\begin{align}
	\left\langle u_{\mathbf{k}}^{v}\left|\left[H,x+iy\right]\right|u_{\mathbf{k}}^{c}\right\rangle & \propto\mathcal{A}^{\gamma_{0}}_{c,v}\left(\tau + 1\right)e^{-2i \tau \theta}+\nonumber\\
	& \quad +\mathcal{A}^{\gamma_{0}}_{v,c}\left(\tau - 1\right)e^{-4i \tau \theta}.\label{eq:circ_g0}
\end{align} 
The $\left(\tau + 1\right),\,\left(\tau - 1\right)$ factors further restrict the selection rules, only allowing those to $ d_{+} $--series states in the $\tau=1$ valley and those to $ g_{-} $--series states in the $\tau=-1$ valley. 

\subsection{Trigonal Warping Effects\label{sec:trigonal_warp}}

Finally, we will consider the effects of the additional hopping parameters which were discarded when solving the Bethe--Salpeter equation, namely $\gamma_{3}$ and $\gamma_{4}$. Although $ \gamma_{2} $ was also discarded, this hopping parameter will not contribute to the optical selection rules as it appears in the tight binding Hamiltonian as a constant term. Computing the dipole operator matrix element while considering the $ \gamma_{3} $ hopping term results in two new allowed transitions. 

For $x$--aligned linearly polarized light, terms proportional to $ \gamma_{3} $ lead to 
\begin{align}
	\left.\left\langle u_{\mathbf{k}}^{v}\left|\left[H,x\right]\right|u_{\mathbf{k}}^{c}\right\rangle\right|_{\gamma_3} & \propto\mathcal{B}^{\gamma_{3}}_{c,v}e^{-i \tau \theta} +\mathcal{B}^{\gamma_{3}}_{v,c}e^{-5i \tau \theta},
\end{align}
with 
\begin{align*}
	\mathcal{B}^{\gamma_{3}}_{c,v}&=\psi_{1,c}\psi_{4,v}+\psi_{3,c}\psi_{6,v},\\
	\mathcal{B}^{\gamma_{3}}_{v,c}&=\psi_{1,v}\psi_{4,c}+\psi_{3,v}\psi_{6,c},
\end{align*}
only allowing transitions to $ m=\tau $ ($ p $ series) and $ m=5\tau $ ($ h $ series) states. Comparing the relative amplitudes of the contributions from both series, those from $p_{+}$--series states dominate and $h_{+}$--series states go totally unnoticed, with relative amplitudes again less than $ 0.1\% $.
Terms proportional to $ \gamma_{4} $, in turn, lead to 
\begin{align}
	\left.\left\langle u_{\mathbf{k}}^{v}\left|\left[H,x\right]\right|u_{\mathbf{k}}^{c}\right\rangle\right|_{\gamma_4} & \propto\mathcal{C}^{\gamma_{4}}_{c,v}e^{-2i \tau \theta}+\mathcal{C}^{\gamma_{4}}_{v,c}e^{-4i \tau \theta},
\end{align} 
with 
\begin{align*}
	\mathcal{C}^{\gamma_{4}}_{c,v}&=\psi_{1,c}\psi_{3,v}+\psi_{2,c}\psi_{4,v}+\psi_{3,c}\psi_{5,v}+\psi_{4,c}\psi_{6,v},\\
	\mathcal{C}^{\gamma_{4}}_{v,c}&=\psi_{1,v}\psi_{3,c}+\psi_{2,v}\psi_{4,c}+\psi_{3,v}\psi_{5,c}+\psi_{4,v}\psi_{6,c},
\end{align*}
imposing the same selection rules obtained for $ \gamma_{0} $. As such, and as $ \gamma_{0}\gg\gamma_{4} $, these will be ignored when computing the optical conductivity.

For right circularly polarized light a similar valley--dependent selection rule to the one obtained in Eq. (\ref{eq:circ_g0}) is present. Explicitly expanding the terms proportional to $ \gamma_{3} $ in the commutator, we obtain 
\begin{align}
	\left.\left\langle u_{\mathbf{k}}^{v}\left|\left[H,x+iy\right]\right|u_{\mathbf{k}}^{c}\right\rangle\right|_{\gamma_3} & \propto\mathcal{B}^{\gamma_{3}}_{c,v}\left(\tau - 1\right)e^{-i \tau \theta}+\nonumber\\
	&\quad+\mathcal{B}^{\gamma_{3}}_{v,c}\left(\tau + 1\right)e^{-5 i \tau \theta},
\end{align} 
allowing only transitions to $ m=5\tau $ ($ h $ series) in the $ \tau=1 $ valley and to $ m=\tau $ ($ p $ series) in the $ \tau=-1 $ valley.
Analogously to linearly polarized light, terms proportional to $ \gamma_{4} $ lead to 
\begin{align}	
	\left.\left\langle u_{\mathbf{k}}^{v}\left|\left[H,x+iy\right]\right|u_{\mathbf{k}}^{c}\right\rangle\right|_{\gamma_4} & \propto\mathcal{C}^{\gamma_{4}}_{c,v}\left(\tau+1\right)e^{-2i \tau \theta}+\nonumber\\
	&\quad +\mathcal{C}^{\gamma_{4}}_{v,c}\left(\tau-1\right)e^{-4i \tau \theta},
\end{align} 
allowing only transitions to $ m=2\tau $ ($ d $ series) in the $ \tau=1 $ valley and to $ m=4\tau $ ($ g $ series) in the $ \tau=-1 $ valley. 

The contribution to the optical conductivity from trigonal warping goes mostly unnoticed as the intensity is close to two orders of magnitude smaller $ \left(\gamma_{3}^2/\gamma_{0}^2\approx0.01\right) $, and the only distinguishable transition is that which is associated with the $ 2p_{+}$ resonance. This occurs as this resonance is much larger than all other $ p $--series resonances and occurs far enough from the resonances originating from the dominant hopping parameter $ \gamma_{0} $. 

\begin{figure*}
	\centering
	\includegraphics{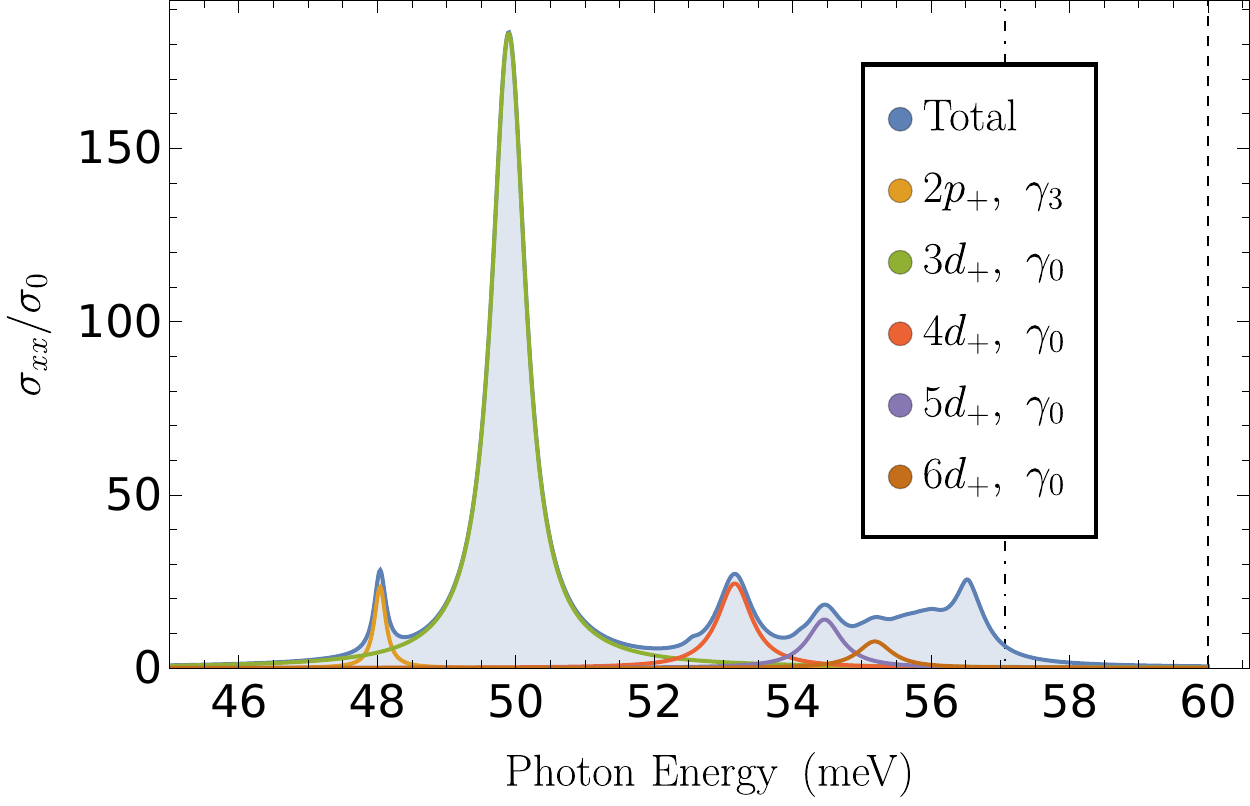}
	
	\caption{Real part of the excitonic $xx$--conductivity for biased ABC--stacked
		trilayer graphene encapsulated in hBN with a bias potential $V=30\,\mathrm{meV}$,
		broadening parameters $\Gamma_{nd_+}=0.3\,\mathrm{meV}$ and $\Gamma_{np_+}=0.1\,\mathrm{meV}$, and a $N=450$
		point Gauss--Legendre quadrature. First ten states of each excitonic series were considered
		for the total conductivity. Vertical
		dashed lines represent the bandgap at $k=0$ (right) and at the band extremes (left). The different $ \gamma $s in the legend symbolize the hopping term that leads to specific resonances. \label{fig:conductivity_xx_full}}
\end{figure*}

Computing the sum over all the previously mentioned states, with $ 10 $ states for each allowed transition, we plot the real part of the $ xx $-linear optical conductivity in Fig. \ref{fig:conductivity_xx_full} for a external bias of $V=30\,\mathrm{meV}$. The first few states contributing to the optical conductivity are also plotted individually as to clearly identify each resonance and they are labeled according to the hopping parameter that allows the transition in question. In this figure, we can clearly distinguish three resonances, namely those associated with $ 2p_+ $, $ 3d_+ $, and $ 4d_+ $ states, with a plateau forming close to the bandgap value as the excitonic resonances become ever closer to each other. The location and amplitude of these resonances are extremely sensitive to the external bias, as we will now see in Sec. \ref{sec:tuning}.
 
\subsection{Tunability via Bias Potential}\label{sec:tuning}

To conclude our study of the ABC-trilayer graphene optical conductivity, we will now analyze the tunability of the excitonic resonances via the bias potential, considering a broad range of external biases and computing the excitonic conductivity for the systems in question. It is important to note that changing the bias potential will also alter the effective screening length present in the Rytova--Keldysh potential (as discussed in Appendix \ref{app:Effective-Screening-Length}) and we will therefore need to recompute the effective screening length for each individual external bias. Additionally, it is also important to note that, as was discussed in Sec. \ref{sec:Bethe-Salpeter-Equation}, the lowest energy bands only dominate the low energy response of the system for sufficiently low external biases. As such, we only compute the excitonic optical conductivity for external biases up to $ V=110\,\mathrm{meV} $. At this external bias, the contributions from higher bands to $ \boldsymbol{\Omega}_{n,\alpha} $ (Eq. (\ref{eq:tg_pedersen_OMEGA})) are still negligible, further justifying the use of only the two bands closest to the gap in our calculations.

\begin{figure*}
	\centering	\includegraphics{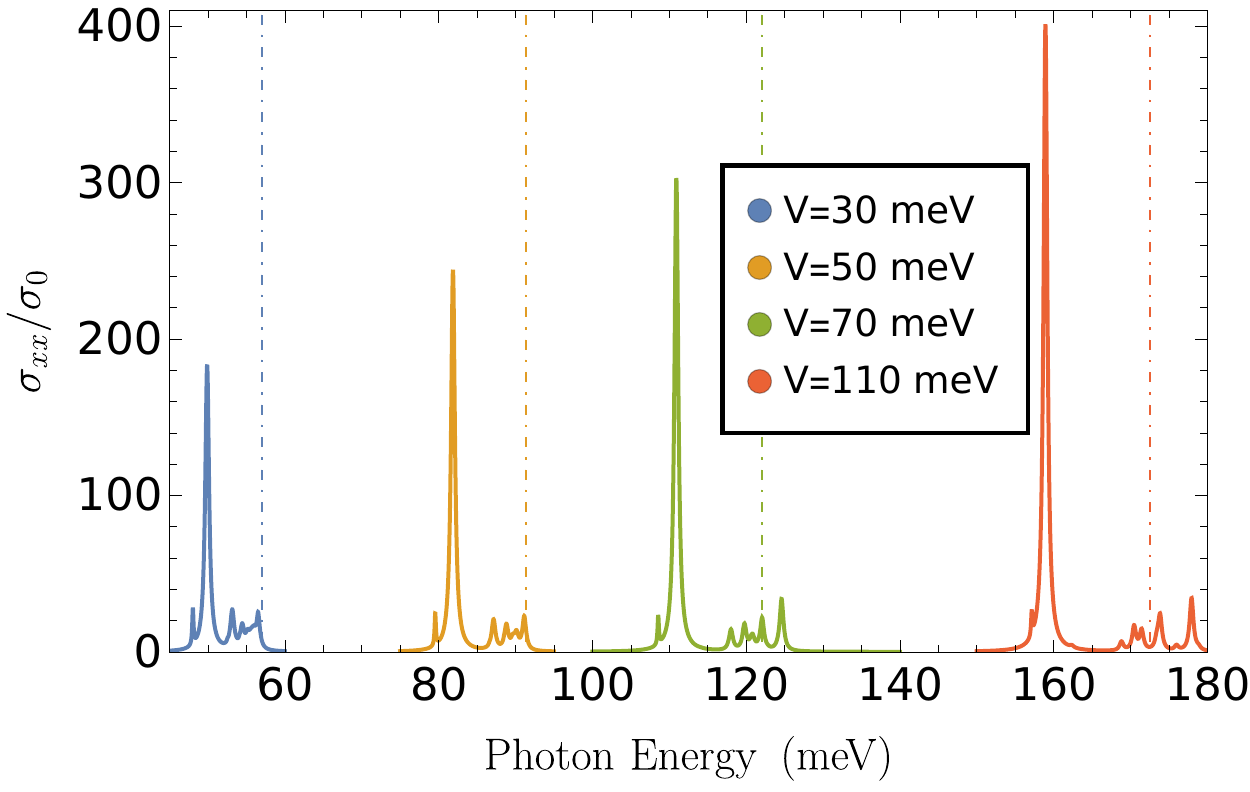}
		
	\caption{Real part of the excitonic $xx$--conductivity for biased ABC--stacked
	trilayer graphene encapsulated in hBN with various bias potentials $V=30$, $50$, $70$, and $110\,\mathrm{meV}$,
	broadening parameters $\Gamma_{nd_+}=0.3\,\mathrm{meV}$ and $\Gamma_{np_+}=0.1\,\mathrm{meV}$, and a $N=450$ point Gauss--Legendre quadrature. First ten states of each excitonic series were considered for the total conductivity. Vertical
	dot--dashed lines represent the bandgap at 
	the band extremes, while the bandgap at $ k=0 $ will be simply $ 2V $. Leftmost resonance in each curve is associated with the transition to the $ 2p_+ $ excitonic state, while the dominant peak and those to its right are associated with the transition to the $ 3d_+ $ and higher $ nd_+ $ excitonic states, respectively.}\label{fig:conductivity_xx_V_multi}
\end{figure*}

The real part of the resulting optical conductivity for various external biases is plotted in the right panel of Fig. \ref{fig:conductivity_xx_V_multi}, together with several dashed lines representing the bandgap characteristic of each system. Analogously to what was discussed in Fig. \ref{fig:conductivity_xx_full}, the optical conductivity plotted in Fig. \ref{fig:conductivity_xx_V_multi} takes into account both the dominant transitions allowed by the $ \gamma_{0} $ hopping and those originating from trigonal warping (modeled by the $ \gamma_{3} $ parameter).

As it can be observed, the relative amplitude of the dominant resonance increases as the external bias increases, leading to it overpowering the nearby $ 2p_+ $ resonance for larger biases (see line for $ V=110\,\mathrm{meV} $). Above $ V=50\,\mathrm{meV} $, resonances associated with higher $ nd_+ $ begin to appear at energies above the dashed lines of the bandgap. These states are, however, still well within the $ \Delta=2V $ gap at $ k=0 $ and, upon inspection of their density plots (similarly to Fig. \ref{fig:density_plot_V_multi}), these appear to be getting more localized near $ k=0 $, implying higher delocalization in real space, as one would expect from higher energy states. 

\section{Conclusion}
In this paper we studied the excitonic optical response of biased rhombohedral trilayer graphene.To this end, we began by reviewing the single particle electronic properties of the multilayer system by considering a simplified tight binding Hamiltonian. The eigenstates of this tight binding Hamiltonian are then used as the input states for the Bethe--Salpeter equation, whose solution leads to the excitonic states. 

With the excitonic wave functions and binding energies known, we proceeded to the computation of the optical conductivity of the trilayer. This allowed us to study the optical selection rules for excitonic transitions while also giving valuable insight into the strength of the photon--exciton coupling. We found that, if trigonal warping is ignored, only $ d $ and $ g $--series states are optically bright, although the oscillator strength for $ g $--states is negligible when compared to that of $ d $--states.
When trigonal warping is taken into account, new transitions become optically bright, as was expected from the symmetry breaking this new hopping parameter introduces. The new couplings make both $ p $ and $ h $--series states optically bright, although the contribution from $ h $--series states is again negligible. Additionally, due to the small amplitude of the trigonal warping parameter relative to the dominant hopping term, only the $ 2p $ state presents a relevant contribution to the optical conductivity. 
Additional hopping parameters were also studied, namely hopping terms between same sublattice sites on different layers. The optical selection rules generated were identical to those from the dominant hopping term, allowing us to discard this contribution due to the much smaller hopping parameter. 

Varying the external bias potential, we observed an increase in the localization of the exciton as the bias increases, with the state associated with the dominant excitonic resonance spread about $ 80\, \text{\AA}$ in real space at an external bias of $ 110\,\mathrm{meV} $. We also observed that the relative amplitude of the dominant excitonic resonance, associated with the $ 3d $ excitonic state, increased as the potential increases. The smaller $ 2p $ resonance becomes increasingly masked by its proximity to the dominant peak, becoming almost indistinguishable from the $ 3d $ resonance at an external bias of $ 110\,\mathrm{meV} $.

\section*{Acknowledgments}

M. F. C. M. Q. acknowledges the International Nanotechnology Laboratory (INL) and the Portuguese Foundation for Science and Technology (FCT) for the Quantum Portugal Initiative grant SFRH/BD/151114/2021. 
N. M. R. P. acknowledges support by the Portuguese Foundation for Science and Technology (FCT) in the framework of the Strategic Funding UIDB/04650/2020, COMPETE 2020, PORTUGAL 2020, FEDER, and  FCT through projects POCI-01-0145-FEDER-028114, POCI-01-0145-FEDER-02888 and PTDC/NANOPT/ 29265/2017, PTDC/FIS-MAC/2045/2021, and from the European Commission through the project Graphene Driven Revolutions in ICT and Beyond (Ref. No. 881603, CORE 3).
\begin{widetext}
\appendix
\section{Effective Screening Length \label{app:Effective-Screening-Length}}

The effective screening length is given by \cite{PhysRevB.99.035429}
\begin{equation}
	r_{0}=\frac{\hbar^{3}c\alpha}{\pi m_{0}^{2}}\sum_{c,v}\int\frac{\left|\left\langle u_{\mathbf{k}}^{c}\left|P_{x}\right|u_{\mathbf{k}}^{v}\right\rangle \right|^{2}}{\left[E_{c}\left(k\right)-E_{v}\left(k\right)\right]^{3}}k\,dk\,d\theta.
\end{equation}
Substituting the momentum matrix element, defined as 
\[
P_{x}=\frac{m_{0}}{\hbar}\frac{\partial H}{\partial k_{x}},
\]
we obtain
\begin{equation}
	r_{0}=\frac{\hbar c\alpha}{\pi}\sum_{c,v}\int\frac{\left|\left\langle u_{\mathbf{k}}^{c}\left|\frac{\partial H}{\partial k_{x}}\right|u_{\mathbf{k}}^{v}\right\rangle \right|^{2}}{\left[E_{c}\left(k\right)-E_{v}\left(k\right)\right]^{3}}k\,dk\,d\theta.\label{eq:screen}
\end{equation}
This effective screening length is, as can be seen in Eq. (\ref{eq:screen}, very sensitive to the external bias, falling quickly for higher values of the external bias\cite{PhysRevB.99.035429}.

Considering only the lowest energy bands, dominant for low bias potentials, we obtain
\begin{align}
	r_{0} & =\frac{\hbar c\alpha}{\pi}\int\frac{\left|\left\langle u_{\mathbf{k}}^{c,-1}\left|\frac{\partial H}{\partial k_{x}}\right|u_{\mathbf{k}}^{v,-1}\right\rangle \right|^{2}}{\left[E_{c,-1}\left(k\right)-E_{v,-1}\left(k\right)\right]^{3}}k\,dk\,d\theta.\label{eq:screen_2bands}
\end{align}
For $ V=50\,\mathrm{meV}$, the value of this screening length will be $ r_{0}=165.623\,\text{\AA} $.

\section{Bethe--Salpeter Equation \label{app:BSE}}
Taking the thermodynamic limit, Eq. (\ref{eq:BSE-simplified}) can be written as
\begin{align}
E \, f_{c,\eta_{1};v,\eta_{4}}\left(k\right) & =\left(E_{\mathbf{k}}^{c,\eta_{1}}-E_{\mathbf{k}}^{v,\eta_{4}}\right)f_{c,\eta_{1};v,\eta_{4}}\left(\mathbf{k}\right)-\label{eq:BSE-limit}\\
&-\frac{1}{4\pi^{2}}\sum_{\eta_{2},\eta_{3}}\int qdqd\theta_{q}V\left(\mathbf{k}-\mathbf{q}\right)\left\langle u_{\mathbf{k}}^{c,\eta_{1}}\mid u_{\mathbf{q}}^{c,\eta_{2}}\right\rangle \left\langle u_{\mathbf{q}}^{v,\eta_{3}}\mid u_{\mathbf{k}}^{v,\eta_{4}}\right\rangle f_{c,\eta_{2};v,\eta_{3}}\left(q\right)e^{im\left(\theta_{q}-\theta_{k}\right)}.\nonumber
\end{align}
This problem can be simplified further, as $\left\langle u_{\mathbf{k}}^{c,\eta_{1}}\mid u_{\mathbf{q}}^{c,\eta_{2}}\right\rangle \left\langle u_{\mathbf{q}}^{v,\eta_{3}}\mid u_{\mathbf{k}}^{v,\eta_{4}}\right\rangle $
consists of a sum of different term with well--defined phases if a careful choice of the spinor phases has been made (Eq. (\ref{eq:generic_ev})). For compactness, in this Appendix we will suppress the $ \eta $ indices, instead using $ c,c^{\prime},v,v^{\prime} $ to distinguish the different bands which take part in the calculation. 
As such, it can be written as 
\begin{align}
	&	\left\langle u_{\mathbf{k}}^{c}\mid u_{\mathbf{q}}^{c^{\prime}}\right\rangle \left\langle u_{\mathbf{q}}^{v^{\prime}}\mid u_{\mathbf{k}}^{v}\right\rangle=\sum_{\lambda}\mathcal{A}_{\lambda}^{cc^{\prime}vv^{\prime}}\left(k,q\right)e^{i\lambda\left(\theta_{q}-\theta_{k}\right)},
\end{align}
where the angular dependence has been extracted from $ \mathcal{A}_{\lambda}^{cc^{\prime}vv^{\prime}}\left(k,q\right) $. 

Regarding the radial integral of the potential term, it can be written
as 
\begin{equation}
	I_{m}\left(k,q\right)=\int_{0}^{2\pi}\frac{\cos\left(m\theta\right)}{\kappa\left(k,q,\theta\right)\left[1+r_{0}\kappa\left(k,q,\theta\right)\right]}d\theta,
\end{equation}
where $\kappa\left(k,q,\theta\right)=\sqrt{k^{2}+q^{2}-2kq\cos\left(\theta\right)}$
and only the even term is non--zero due to parity. Inspecting the
integrand, it is clear that the $I$ function will be numerically
ill--behaved when $k=q$. For this effect, we decompose the integrand
in terms of partial functions as 
\begin{align}
	I_{m}\left(k,q\right) & =\int_{0}^{2\pi}\frac{\cos\left(m\theta\right)}{\kappa\left(k,q,\theta\right)}d\theta-r_{0}\int_{0}^{2\pi}\frac{\cos\left(m\theta\right)}{1+r_{0}\kappa\left(k,q,\theta\right)}d\theta\nonumber\\
	& =J_{m}\left(k,q\right)-K_{m}\left(k,q\right).\nonumber 
\end{align}
With this decomposition, it is clear now that only the $J_{m}\left(k,q\right)$
integral will be problematic when $k=q$. Substituting $I_{m}\left(k,q\right)$
into Eq. (\ref{eq:BSE-limit}), we write 
\begin{align}
	& Ef_{cv}\left(k\right)=\left(E_{k}^{c}-E_{k}^{v}\right)f_{cv}\left(k\right)-\nonumber \\
	& -\frac{1}{4\pi^{2}}\sum_{c^{\prime}v^{\prime}}\int_{0}^{+\infty}\sum_{\lambda}\left\{ J_{m+\lambda}\left(k,q\right)\mathcal{A}_{\lambda}^{cc^{\prime}vv^{\prime}}\left(k,q\right)-K_{m+\lambda}\left(k,q\right)\mathcal{A}_{\lambda}^{cc^{\prime}vv^{\prime}}\left(k,q\right)\right\} f_{c^{\prime}v^{\prime}}\left(q\right)qdq
\end{align}

Writing
\begin{align*}
\mathcal{J}_{m}^{cc^{\prime}vv^{\prime}}\left(k,q\right)&=\sum_{\lambda}J_{m+\lambda}\left(k,q\right)\mathcal{A}_{\lambda}^{cc^{\prime}vv^{\prime}}\left(k,q\right),
 & \mathcal{K}_{m}^{cc^{\prime}vv^{\prime}}\left(k,q\right)&=\sum_{\lambda}K_{m+\lambda}\left(k,q\right)\mathcal{A}_{\lambda}^{cc^{\prime}vv^{\prime}}\left(k,q\right),
\end{align*}
the BSE can now be compactly written as 
\begin{align}
	Ef_{cv}\left(k\right)&=\left(E_{k}^{c}-E_{k}^{v}\right)f_{cv}\left(k\right)-\frac{1}{4\pi^{2}}\sum_{c^{\prime}v^{\prime}}\int_{0}^{+\infty}\left[ \mathcal{J}_{m}^{cc^{\prime}vv^{\prime}}\left(k,q\right)-\mathcal{K}_{m}^{cc^{\prime}vv^{\prime}}\left(k,q\right)\right] f_{c^{\prime}v^{\prime}}\left(q\right)qdq.\label{eq:final_BSE}
\end{align}

We now focus our attention on the problematic $\mathcal{J}_{m}^{cc^{\prime}vv^{\prime}}\left(k,q\right)$
object. To treat the divergence at $q=k$, an auxiliary function $g_{m}\left(k,q\right)$
is introduced. This function obeys the limit 
\[
\lim_{q\rightarrow k}\left[\mathcal{J}_{m}^{cc^{\prime}vv^{\prime}}\left(k,q\right)-g_{m}\left(k,q\right)\right]=0
\]
and it modifies the integrals as
\begin{align}
	\int_{0}^{+\infty}\mathcal{J}_{m}^{cc^{\prime}vv^{\prime}}\left(k,q\right)f_{c^{\prime}v^{\prime}}\left(q\right)qdq\rightarrow & \int_{0}^{+\infty}\left[\mathcal{J}_{m}^{cc^{\prime}vv^{\prime}}\left(k,q\right)-g_{m}\left(k,q\right)\right]f_{c^{\prime}v^{\prime}}\left(q\right)qdq+\nonumber\\
	&\qquad\qquad+f_{c^{\prime}v^{\prime}}\left(k\right)\int_{0}^{+\infty}g_{m}\left(k,q\right)qdq.
\end{align}
Following \cite{PhysRevB.43.6530,PhysRevB.105.045411}, this auxiliary function is chosen as 
\[
g_{m}\left(k,q\right)=\mathcal{J}_{m}^{cc^{\prime}vv^{\prime}}\left(k,q\right)\frac{2k^{2}}{k^{2}+q^{2}}.
\]

Having finished outlining the analytical procedure, we now proceed
to the numerical solution of the BSE. This is performed using the same methodology as \cite{PhysRevB.105.045411}, which we will quickly outline. 
A variable change is introduced
as to convert the integration limits from $\left[0,+\infty\right)$
to a finite limit, in this case $\left[0,1\right]$, defined as $q=\tan\left(\frac{\pi x}{2}\right)$. With this variable change, we proceed by discretizing $x$, writing the numeric
problem as 
\begin{align} 
	& Ef_{cv}\left(k_{i}\right)=\left(E_{k_{i}}^{c}-E_{k_{i}}^{v}\right)f_{cv}\left(k_{i}\right)+\frac{1}{4\pi^{2}}\sum_{c^{\prime}v^{\prime}}\sum_{j=1}^{N}\left[\mathcal{K}_{m}^{cc^{\prime}vv^{\prime}}\left(k_{i},q_{j}\right)f_{c^{\prime}v^{\prime}}\left(q_{j}\right)q_{j}\frac{dq}{dx_{j}}\right]-\\
		& -\frac{1}{4\pi^{2}}\sum_{c^{\prime}v^{\prime}}\left\{ \sum_{j\neq i}\left[\mathcal{J}_{m}^{cc^{\prime}vv^{\prime}}\left(k_{i},q_{j}\right)f_{c^{\prime}v^{\prime}}\left(q_{j}\right)+g_{m}\left(k_{i},q_{j}\right)\right]q_{j}\frac{dq}{dx_{j}}w_{j}-f_{c^{\prime}v^{\prime}}\left(k_{i}\right)\int_{0}^{\infty}g_{m}\left(k_{i},p\right)pdp\right\},\nonumber\label{eq:BSE}
\end{align}
where $N$ is the number of points considered in the discretization, $w$ is the weight function of the quadrature in question, and the discretized variables are defined as $q_{i}\equiv q\left(x_{i}\right)$,
and $\frac{dq}{dx_{i}}\equiv\left.\frac{dq}{dx}\right|_{x=x_{i}}$.
It is important to note that, while $\int_{0}^{\infty}\mathcal{J}_{m}^{cc^{\prime}vv^{\prime}}\left(k,q\right)qdq$
is numerically problematic at $ q=k $, $\int_{0}^{\infty}g_{m}\left(k,q\right)qdq$
is well--behaved.

In this paper, we employ a Gauss--Legendre quadrature \cite{Kythe2002}, defined
as 
\[
\int_{a}^{b}f\left(x\right)dx\approx\sum_{i=1}^{N}f\left(x_{i}\right)w_{i},
\]
where 
\[
x_{i}=\frac{a+b+\left(b-a\right)\xi_{i}}{2}
\]
with $\xi_{i}$ the $i$-th zero of the Legendre polynomial $P_{N}\left(x\right)$,
and 
\[
w_{i}=\frac{b-a}{\left(1-\xi_{i}^{2}\right)\left[\left.\frac{dP_{N}\left(x\right)}{dx}\right|_{x=\xi_{i}}\right]^{2}}.
\]

Finally, it is important to realize that Eq. (\ref{eq:final_BSE}) can be
written as the eigenvalue problem of a $9N\times9N$ matrix (\emph{i.e.},
a $9\times9$ matrix of $N\times N$ matrices). The $81$ blocks come from the different
combinations of band indices, and each $N\times N$ matrix comes from the numerical discretization of the integral. Solving this eigenvalue problem for a sufficiently large quadrature, one obtains the excitonic eigenvalues and eigenfunctions. 
\end{widetext}

\bibliographystyle{unsrt}

\end{document}